\documentclass[aps,onecolumn,11pt,floatfix,altaffilletter,tightenlines,showpacs,showkeys,notitlepage,nofootinbib,preprintnumbers]{revtex4-1}
\pdfoutput=1

\usepackage{epsfig}
\usepackage{amsmath}
\usepackage{amssymb}
\usepackage{slashed}
\usepackage{subfigure}
\usepackage{cancel}
\usepackage{textcomp}
\usepackage{calc}
\usepackage{graphicx}
\usepackage{dcolumn}
\usepackage{color}
\usepackage[utf8]{inputenc}

\usepackage[colorlinks=true,citecolor=blue,linkcolor=blue]{hyperref}
\usepackage[capitalize]{cleveref}

\allowdisplaybreaks

\catcode`\@=11
\def\lsim{\mathrel{\mathpalette\@versim<}}
\def\gsim{\mathrel{\mathpalette\@versim>}}
\def\@versim#1#2{\vcenter{\offinterlineskip
\ialign{$\m@th#1\hfil##\hfil$\crcr#2\crcr\sim\crcr } }}
\catcode`\@=12

\usepackage{xcolor}

\newcommand{\Slash}[1]{{\ooalign{\hfil/\hfil\crcr$#1$}}}

\newcommand{\al}[1]{\begin{align}#1\end{align}}
\newcommand{\bp}{\begin{pmatrix}}
\newcommand{\ep}{\end{pmatrix}}
\newcommand{\nn}{\nonumber}

\newcommand{\bs}[1]{\boldsymbol}

\newcommand{\be}{\begin{eqnarray}}
\newcommand{\ee}{\end{eqnarray}}


\definecolor{rubinered}{rgb}{0.82,0.0,0.34}
\definecolor{upforestgreen}{rgb}{0.0,0.27,0.13}
\definecolor{burntorange}{rgb}{0.8,0.33,0.0}

\begin{document}


\title{Heavy Dark Matter, Neutrino Masses and Higgs Naturalness \\
from a Strongly Interacting Hidden Sector}

\author{Mayumi Aoki$^{1}$}   \email{mayumi.aoki@staff.kanazawa-u.ac.jp}
\author{Vedran Brdar$^{2}$} \email{vbrdar@mpi-hd.mpg.de}
\author{Jisuke Kubo$^{2,3\,}$} \email{kubo@mpi-hd.mpg.de}
\affiliation{
$^1$Institute  for  Theoretical  Physics,  Kanazawa  University,  Kanazawa  920-1192,  Japan\\
$^2$Max-Planck-Institut f\"ur Kernphysik,
       69117~Heidelberg, Germany\\
$^3$Department  of  Physics,  University  of  Toyama,  3190  Gofuku,  Toyama  930-8555,  Japan}

\preprint{KANAZAWA-20-05}
\vspace*{1cm}

\begin{abstract}

We consider the extension of the Standard Model (SM) with 
a strongly interacting QCD-like hidden sector, at least two generations of right-handed neutrinos and one scalar singlet. Once scalar singlet obtains a nonzero vacuum expectation value, active neutrino masses are generated through type-I seesaw mechanism. Simultaneously, the electroweak scale is generated through the radiative corrections involving these massive fermions. This is the essence of the scenario that is known as the ``neutrino option" for which the successful masses of right-handed neutrinos are in the range $10^7-10^8$ GeV. The main goal of this work is to scrutinize the potential to accommodate dark matter in such a realization. The dark matter candidates are Nambu-Goldstone bosons which appear due to the dynamical breaking of the hidden chiral symmetry. The mass spectrum studied in this work is such that masses of Nambu-Goldstone  bosons and singlet scalar exceed those of right-handed neutrinos. Having the masses of all relevant particles several orders of magnitude above $\mathcal{O}$(TeV),
the freeze-out of dark matter is not achievable and hence we turn to alternative scenarios, namely freeze-in. 
The Nambu-Goldstone bosons can interact with particles that are not in SM but, however, have non-negligible abundance through their not-too-small couplings with SM. Utilizing this, we demonstrate that the dark matter in the model is successfully produced at temperature scale where the right-handed neutrinos are still stable. We note that the lepton number asymmetry sufficient for the generation of observable baryon asymmetry of the Universe can be produced in right-handed neutrino decays. Hence, we infer that the model has the potential to simultaneously address several of the most relevant puzzles in contemporary high-energy physics.

\end{abstract}

\maketitle
\newpage 

\section{Introduction}
\label{sec:intro}
\noindent
Despite being a great success, the Standard Model (SM) has several shortcomings. It predicts all three species of neutrinos to be massless, in contrast to the observation from neutrino oscillation experiments. It also does not contain a particle that can be a viable dark matter (DM) candidate as well as a successful mechanism for generation of baryon asymmetry of the Universe. Many of the proposed theories Beyond the Standard Model (BSM) predict degrees of freedom at very high scales which could for instance address aforementioned issues. However, such an approach has its own difficulties: increment of the mass scale at which heavy particles reside leads to the 
larger Higgs mass correction through processes at loop level, and this leads to a feature dubbed hierarchy problem. In this paper we will try to address
most of the aforementioned issues simultaneously, within a framework that is a well-motivated SM extension. The core is connection between generation of light neutrino masses and electroweak scale via processes involving right handed neutrinos. 
The light neutrino masses induced via type-I seesaw 
\cite{Minkowski,GellMann:1980vs,Yanagida:1979as,Goran}
are 
\begin{align}
m_\nu \simeq \frac{y_\nu^2 v_h^2}{m_N}\,,\,
\label{eq:nu_masses}
\end{align}
where $y_\nu$ represents lepton portal Yukawa coupling, $v_h=246$ GeV and $m_N$ stands for right-handed neutrino mass.
If the right-handed neutrinos are heavy, they can significantly contribute
to the correction to the Higgs mass term, ${\cal L}_\text{SM}
\subseteq -\mu_H^2 H^\dag H$, in higher orders in perturbation theory,
where $H$ is the SM Higgs doublet.
At one-loop the  correction (finite term)  is given by 
\cite{Vissani:1997ys,Casas:1999cd,Clarke:2015gwa,Bambhaniya:2016rbb}
\begin{align}
|\Delta \mu_H^2| \sim \frac{y_\nu^2 m_N^2}{4\pi^2}\,,
\label{eq:higgs}
\end{align}
where $\sim$ stands to indicate that the correction depends on the renormalization
scale  at which it is evaluated.
The idea of  the ``neutrino option"
\cite{Brivio:2017dfq}  is based on the assumption that
(\ref{eq:higgs}) is the main source
for the Higgs mass term, i.e. $\mu_H^2 \simeq\Delta \mu_H^2$.
Simultaneous solution of (\ref{eq:nu_masses}) and $\mu_H^2 \simeq\Delta \mu_H^2$
reveals $m_N$ between $10$ and $100$ PeV, with $y_\nu\lesssim 10^{-4}$
\cite{Brivio:2017dfq} (see also Refs.\,\cite{Brivio:2018rzm,Davoudiasl:2014pya}). 
The neutrino option thus establishes a link between
the heavy  right-handed neutrinos and the electroweak scale. 

We are naturally led to the desire to
 embed  the neutrino option into a classically scale invariant theory
 \cite{Bardeen:1995kv},
because the Higgs mass term, which is the only 
scale symmetry violating term  in the SM Lagrangian, is assumed to be  absent or extremely suppressed for the neutrino option to be sensible:
We would like to understand the origin of the electroweak scale
as a consequence of spontaneous (dynamical) scale symmetry breaking
\cite{PhysRevD.7.1888,Hempfling:1996ht,Meissner:2006zh}.
To be complete, we also have to understand 
the origin of the right-handed neutrino mass $m_N$
in this manner.
A simplest way is to realize $m_N\sim \langle S \rangle$
\cite{Meissner:2006zh},
where $S$ stands for  a SM singlet real scalar field.
That is, the Majorana mass term 
for the  right-handed neutrino $N_R$
is replaced by the Yukawa 
interaction between $S$ and  $N_R$ \cite{Meissner:2006zh}.
Along this line, a conformal UV completion of  the neutrino option
has been performed in Ref.\,\cite{Brdar:2018vjq}:
The  mass squared correction
 $\Delta \mu_H^2$ transmutes indeed into
radiative correction to the dimensionless coupling 
$\lambda_{HS}$ before the spontaneous scale symmetry 
 breaking  \cite{Brdar:2018vjq}, where $\lambda_{HS}$ is 
 the coupling  for $S^2 H^\dag H$,  and 
 it is assumed that its one-loop correction, $\sim
 y_\nu^2 y_M^2/16\pi^2$, is
 the main source for this quartic coupling.
 
The main goal of this work is to successfully embed DM while preserving the  aforementioned neutrino option property. To this end, we will introduce a strongly interacting QCD-like hidden sector, 
in which the mass scale is generated in a non-perturbative way via condensation that breaks chiral symmetry dynamically
\cite{Nambu:1960xd,Nambu:1961tp,Nambu:1961fr}.
Such scenario is in contrast with the aforementioned  conformal UV completion of the neutrino option \cite{Brdar:2018vjq,Brdar:2018num}, where the scale is generated perturbatively \`{a} la Coleman-Weinberg \cite{PhysRevD.7.1888}. 
Realistic models  employing  QCD-like strong dynamics have 
been proposed
in Refs.\,\cite{Hur:2011sv,Heikinheimo:2013fta,Holthausen:2013ota,Kubo:2014ova,Kubo:2015cna,Hatanaka:2016rek}.
A crucial observation in the presence of QCD-like hidden sector 
is that it features the appearance of (quasi) Nambu-Goldstone (NG) bosons,
hidden mesons, arising through dynamical chiral symmetry breaking and
that due to their stability they are good DM candidates
\cite{Hur:2011sv,Heikinheimo:2013fta,Holthausen:2013ota,Hatanaka:2016rek,Kubo:2014ida,Ametani:2015jla}.

We would like to bring  up at this stage that the  use of  (\ref{eq:higgs})
 as the main source
for the Higgs mass term in a scotogenic model,
coupled  with  a QCD-like hidden sector,
has been proposed 
in Ref.\,\cite{Davoudiasl:2014pya}, where the DM candidate sits in an additionally introduced inert Higgs doublet \cite{Barbieri:2006dq}.
Further, the authors of Ref.\,\cite{Samanta:2020gdw} have recently 
shown that the neutrino option can be related to both DM and baryon asymmetry of the universe through inflation. In this work we will take a complementary path and demonstrate successful embedding of the hidden mesons as DM in the model that does not rely on inflaton decay, although features related to primordial physics will be discussed primarily in the context of the
QCD-like hidden sector that we expect to be rather cold.
This is a new aspect compared to the previous models considered in 
Refs.\,\cite{Hur:2011sv,Heikinheimo:2013fta,Holthausen:2013ota,Kubo:2014ova,Kubo:2015cna,Hatanaka:2016rek,Kubo:2014ida,Ametani:2015jla}.

While we will not study leptogenesis 
\cite{Fukugita:1986hr,Buchmuller:2004nz} for the purpose of generating successful baryon asymmetry of the Universe, we point out that our DM production is fully consistent with resonant leptogenesis 
\cite{Flanz:1994yx,Flanz:1996fb,Pilaftsis:1997dr}
that was previously shown to be successful for the neutrino option \cite{Brdar:2019iem,Brivio:2019hrj}. Namely,  DM production precedes right-handed neutrino decays, and in addition any relevant washout processes involving particles from a hidden sector are suppressed by heavy mass scale.

The paper is organized as follows. In \cref{sec:model},
we introduce the set of considered BSM fields and interactions.
In \cref{subsec:NJL} we present techniques employed for the analysis of the strongly interacting sector, namely Nambu-Jona-Lasinio (NJL) description. Further, in \cref{subsec:mass_spec,subsec:DM-couple} we elaborate on
mass spectrum and DM interactions, respectively. \cref{sec:freeze-in-DM} is reserved for our DM analysis: We start by discussing the thermodynamics of the
hidden sector and then move on toward relevant Boltzmann equations and calculations of thermally averaged quantities (\cref{subsec:boltzmann}). Finally, in \cref{subsec:results} we demonstrate success of DM analysis and present results 
 chiefly in the form of numerical scans. In \cref{sec:summary} we conclude.


\section{The model} 
\label{sec:model}
\noindent
As we have announced in \cref{sec:intro}, we consider the model that is scale-invariant at high energies. Instead of Coleman-Weinberg mechanism \cite{PhysRevD.7.1888} for symmetry breaking we employ a strongly-interacting, QCD-like hidden sector to generate a mass scale
\cite{Hur:2011sv, Heikinheimo:2013fta, Holthausen:2013ota, Kubo:2014ida,  Hatanaka:2016rek,Ishida:2016fbp,Haba:2015qbz}. In contrast to the realization in Ref.\,\cite{Brdar:2018vjq} we introduce only one singlet scalar, $S$, which is coupled to right-handed neutrinos, but also interacts with the vector-like hidden fermions $\psi_i~(i=1,\dots,n_f)$ belonging to the fundamental representation
of  $SU(n_c)_H$. The relevant part of the Lagrangian reads
\al{
	\mathcal{L} \supseteq{}& \tfrac{1}{2}\partial_\mu S \partial^\mu S
	+ \tfrac{i}{2} \bar{N}_R \slashed{\partial} N_R 
	-\tfrac{1}{2}\mbox{Tr}~F^2+
\mbox{Tr}~\bar{\psi}(i\gamma^\mu \partial_\mu +
g_H \gamma^\mu G_\mu -
{\bf y} S)\psi \nn\\ 
	&
	- \tfrac{1}{2} y_M S N^T_R C N_R 
	-\left( y_\nu \bar{L} \tilde{H} \,\tfrac{1}{2}(1+\gamma_5)
	N_R + \text{h.c.} \right)
- V_\text{tree}(H,S)\,.
	\label{eq:UVlagn}
	}
	
Here, $H$ ($\tilde{H}=i\sigma_2 H^*$) 
and $L$  are the SM Higgs and lepton doublet, 
$N_R$ denotes the right-handed  neutrinos
($C \bar{N}_R^T=N_R$), whereas  $G_\mu$ is the gauge field 
for the hidden QCD ($F^{\mu\nu}$ denotes corresponding field strength tensor). 
The generation indices of the SM sector  are suppressed in (\ref{eq:UVlagn}).
Strictly speaking, the Yukawa couplings $y_\nu$ and $y_M$ should be
matrices in the generation space. However, we will not consider  the flavor structure but instead
$y_\nu$ and $y_M$ 
will be representative real numbers. This is justifiable 
because the detailed matrix structure of these couplings  is not relevant for dark matter analysis that will be performed in this work.
The  ${\bf y}$ is an $n_f\times n_f$ Yukawa coupling matrix
 which can be taken as a diagonal matrix without loss of generality, 
\emph{i.e.} ${\bf y}=\mbox{diag.}(y_1,\dots,y_{n_f})$, where
the diagonal entries  $y_i$ are assumed to be positive.
The tree-level scalar potential $V_\text{tree}$ in 
(\ref{eq:UVlagn}) reads
\begin{align}
	V_\text{tree}(H,S) =
	\lambda_H (H^\dagger H)^2
	+ \frac{1}{4}\lambda_S S^4
	+\frac{1}{2} \lambda_{HS} S^2 (H^\dagger H)\,,
	\label{eq:UVpotential}
\end{align}
where  $H^T=( H^+,~(h+iG^0)\sqrt{2}  )$ is the SM Higgs doublet field
with $H^+$ and $G^0$ as the would-be NG fields, and
the (tree-level) stability conditions for the scalar potential is given by
\begin{align}
\label{stability}
    \lambda_{H}>0,~~\lambda_{S}>0, ~~2\sqrt{\lambda_{H}\lambda_{S}}+\lambda_{HS}&>0 \,.
\end{align}

In the strongly interacting hidden sector, described by the non-abelian
gauge theory based on $SU(n_c)_H$, 
the $SU(n_c)_H$ invariant chiral bilinear dynamically forms a $U(n_f)$ invariant condensate $\langle \bar{\psi}_i\psi_j \rangle \propto \delta_{ij}$,
and at the same time the real scalar singlet $S$ acquires a vacuum expectation value (VEV), $v_S=
\langle S \rangle$. Sequentially, the right-handed neutrinos become massive with a Majorana mass $m_N=y_M v_S$; this allows for the generation of the loop-induced Higgs mass term that is at the core of the neutrino option idea \cite{Brivio:2017dfq}.

Since the Higgs mass, $m_h$, the VEV of the Higgs field $h$ (denoted as $v_h$) and also
the mass scale  of light neutrinos are experimentally fixed, 
the parameter space in the model is strongly constrained. Namely, as already introduced in \cref{eq:nu_masses,eq:higgs}, $m_h$ and $v_h$ fix the value of the product $y_\nu^2\, m_N^2$,
while the see-saw  mechanism  constrains
$y_\nu^2/m_N^2$.  It turns out that the successful parameter space is between 10 and 100 PeV for $m_N$, while $y_\nu\lesssim 10^{-4}$. Without loss of generality, in our analysis we will stick to the value of $m_N=5\times 10^7$ GeV and the corresponding value of $y_\nu$ with which the scale of light neutrinos is fixed to $m_\nu \simeq 0.1$ eV.

Following Refs.\,\cite{Holthausen:2013ota,Kubo:2014ida,Ametani:2015jla} we consider $n_f=n_c=3$.
In this case, the hidden chiral symmetry 
$\mathrm{SU}(3)_{L}\times\mathrm{SU}(3)_{R}$ is dynamically broken to its diagonal subgroup $SU(3)_V$ by the nonzero chiral condensate,
which implies the existence of eight NG bosons.
Strictly speaking, they are quasi NG bosons, because
the Yukawa coupling ${\bf y} S\bar{\psi}\psi$ breaks
explicitly the chiral symmetry, such that
the NG bosons acquire a mass and as we will show in this work can become cold DM candidate; these fields are stable due to  
 the remnant unbroken $SU(3)_V$ flavor group
or its subgroup (depending on the choice of $y_i$).
To simplify the situation we assume in the following discussion
that $y=y_1=y_2=y_3$ so that the unbroken flavor group is $SU(3)_V$.

Consequently, the free parameters of the model are
\al{
g_H,\, \lambda_S,\, \lambda_{HS},\,  y_M~ \mbox{and}~y\,.
}
The Higgs portal, $\lambda_{HS}$, has to be tiny, because such coupling contributes to the Higgs mass term even at tree-level as $\lambda_{HS} v_S^2$ and the idea is to generate the scale through radiative threshold corrections involving right-handed neutrinos instead of employing scalar sector for that purpose.
We require $\lambda_{HS}v_S^2\lesssim m_h^2$ which reads
\al{\lambda_{HS} \lsim 6\times 10^{-12}\,y_M^2\,,
\label{constHS}
}
for employed right-handed neutrino scale. Instead of setting this coupling around the value given in \cref{constHS} we will conservatively study $\lambda_{HS}\to 0$ limit, or in other words we will not rely on the smallness of this parameter to contribute to the generation of DM abundance.

The relevant system to investigate DM contains the NG boson fields $\phi^a~(a=1,\dots,8)$ which are DM candidates in our model, $S$ and $N_R$. We denote
the mass of $\phi^a$ by $m_\text{DM}$ and 
that of $S$ by $m_S\simeq 3 \lambda_S v_S$. 
The dynamics of the system is of a non-perturbative nature.
In the following subsections we will be using an effective theory,
the Nambu--Jona-Lasinio theory, 
to compute the chiral condensate and  $m_\text{DM}$ and 
also to describe the interactions between $\phi^a$ and $S$. 
The couplings relevant for our DM analysis in \cref{sec:freeze-in-DM} are $y$, $y_M$ and $\lambda_S$.

\subsection{Nambu--Jona-Lasinio description}
\label{subsec:NJL}
\noindent
In order to analyze the strongly interacting hidden sector
described by
\be
{\cal L}_{\rm H}=	-\frac{1}{2}\mbox{Tr}~F^2+
\mbox{Tr}~\bar{\psi}(i\gamma^\mu \partial_\mu +
g_H \gamma^\mu G_\mu -
y S)\psi,~\mbox{with}~y=y_1=y_2=y_3~\mbox{and}~
n_c=n_f=3\,, 
\label{LH}
\ee
we replace the Lagrangian ${\cal L}_{\rm H}$  (\ref{LH}) 
by
  the  Nambu--Jona-Lasinio (NJL) Lagrangian 
 that serves as an effective Lagrangian for the dynamical chiral symmetry breaking
 \cite{Nambu:1960xd,Nambu:1961tp}. It reads
 \begin{align}
{\cal L}_{\rm NJL}&=\mbox{Tr}~\bar{\psi}(i\gamma^\mu\partial_\mu 
-y S)\psi+2\,G~\mbox{Tr} ~\Phi^\dag \Phi
+G_D~(\det \Phi+h.c.)\,,
\label{eq:NJL10}
\end{align}
where
\begin{align}
\Phi_{ij}&= \bar{\psi}_i(1-\gamma_5)\psi_j=
\frac{1}{2}\sum_{a=0}^{8}
\lambda_{ji}^a\, [\,\bar{\psi}\lambda^a(1-\gamma_5)\psi\,]\,,
\end{align}
with $\lambda^0=\sqrt{2/3}~{\bf 1}$ and $\lambda^a (a=1,\dots, 8)$ that are Gell-Mann matrices.
The dimensionful parameters $G$ and $G_D$ have canonical dimensions of $-$2 and $-$5, respectively. 
In order to deal with the nonrenormalizable Lagrangian (\ref{eq:NJL10}) we work in 
the Self-Consistent-Mean-Field (SCMF) approximation
of Refs.\,\cite{Kunihiro:1983ej,Hatsuda:1994pi}.
The mean fields $\sigma$ 
and $\phi_a~(a=0,\dots,8)$ are defined in the ``Bardeen-Cooper-Schrieffer" (BCS) vacuum as
\begin{align}
\label{varphi}
\sigma \,\delta_{ij}=- 4 G\left<\bar{\psi}_i \psi_j \right> \,, ~~~
\phi_a =-2 i G\left<\bar{\psi} \gamma_5 \lambda^a\psi  \right>\,,
\end{align}
where the CP-even mean fields corresponding 
to the non-diagonal elements of $\langle \bar{\psi}_i\psi_j \rangle$ are suppressed; they do not play any role
in out analysis.
The NJL Lagrangian, $\mathcal{L}_{\text{NJL}}$, 
is split into two parts as 
 $\mathcal{L}_{\text{NJL}} =\mathcal{L}_{\text{MFA}}+\mathcal{L}_{I}$,
 where $\mathcal{L}_{I}$ is normal ordered with respect
 to the BCS vacuum, \emph{i.e.} $\langle 0|\mathcal{L}_{I}|0\rangle =0$),
while $\mathcal{L}_{\text{MFA}}$ is computed in the SCMF approximation:
\begin{align}
\nonumber
   \mathcal{L}_{\text{MFA}} = &   
   \mathrm{Tr} ~\bar{\psi}(i\Slash{\partial}-M)\psi -i\mathrm{Tr} ~\bar{\psi}\gamma_5 \phi \psi -\frac{1}{8G}\left( 3\sigma^2+2\sum^8 _{a=1} \phi_a \phi_a \right)  \\
 \label{Hidden SCMFA}
    & +\frac{G_D}{8G^2}\left(  -\mathrm{Tr} ~\bar{\psi} \phi^2 \psi + \sum^8 _{a=1} \phi_a \phi_a \mathrm{Tr} ~\bar{\psi}\psi + i\sigma \mathrm{Tr}~ \bar{\psi}\gamma_5 \phi \psi +\frac{\sigma^3}{2G}+\frac{\sigma}{2G}\sum^8 _{a=1} (\phi_a)^2   \right) 
\end{align}
with $\phi=\sum_{a=1}^8~\phi_a \lambda^a$.
Here
$\phi_0$ has been suppressed\footnote{Due to chiral $U(1)$ anomaly $\phi_0$ is not
a NG boson and is not stable.} and 
 $M$ is given by 
\begin{align}
\label{M}
M(S,\sigma)= \sigma+yS-\frac{G_D}{8G^2}\sigma^2\,.
\end{align}

The one-loop effective potential obtained from 
$\mathcal{L}_{\text{MFA}}$
(\ref{Hidden SCMFA}) 
can be obtained by integrating out the hidden fermions:
\begin{align}
V_{\rm NJL}(S,\sigma)
& = \frac{3}{8G}\sigma^2-
\frac{G_D}{16G^3}\sigma^3
-3n_c I_0(M,\Lambda_H)\,.
\label{eq:Vnjl}
\end{align}
Here the  function $I_0$ is given by
\begin{align}
  I_0(M, \Lambda)= \frac{1}{16\pi^2}\left[ \Lambda^4 \ln \left( 1+\frac{M^2}{\Lambda^2 }\right)-M^4 \ln \left( 1+\frac{\Lambda^2 }{M ^2}\right) + \Lambda^2 M ^2\right] \,,
\end{align}
with a four-dimensional momentum cutoff $\Lambda$;
we denote the cutoff in the hidden sector
by $\Lambda_H$.
For a certain interval of  the dimensionless 
parameters $G^{1/2}\Lambda_H$ and $(-G_D)^{1/5}\Lambda_H$
we have $v_S=\left<\sigma\right>\ne 0$ and $v_\sigma=\left<S \right>\ne 0$
\cite{Holthausen:2013ota,Kubo:2014ida,Ametani:2015jla}.
In view of (\ref{varphi}) it is implied that the dynamics of the hidden sector 
creates a non-vanishing chiral condensate 
$\left<0|\,\bar\psi \psi \,|0\right>\ne 0$. 
The actual value  of $\Lambda_H$ can be indirectly fixed by
the right-handed neutrino mass, because $v_S$ is fixed from  $m_N=y_M v_S$ for
a given $m_N$ and $y_M$:
$\lambda_{HS}$ is no longer an active portal coupling for the neutrino option
and hence does not influence the value of $\Lambda_H$.
Further, one can see that 
the potential $V_{\rm NJL}(S,\sigma)$ is asymmetric in $\sigma$ by inspecting the last term in the NJL Lagrangian (\ref{eq:NJL10}) as well as the constituent mass $M$ (\ref{M}); due to latter chiral phase transition can become of first-order.

It is noted that the mean fields $\sigma$ and $\phi_a$ are non-propagating classical fields at the tree level.
Therefore, their kinetic terms are generated by integrating out the hidden fermions at the one-loop level, which will be seen in Section \ref{Mass spectrum} where 
two point functions are calculated.

The NJL parameters for the hidden QCD sector are obtained by scaling-up the values of
$G, G_D$ and the cutoff $\Lambda$ from QCD hadron physics. 
Following Refs.\,\cite{Holthausen:2013ota,Kubo:2014ida,Ametani:2015jla} we employ the following dimensionless combinations
\begin{align}
 G^{1/2}\Lambda_H=1.82\,,~~~~(-G_D)^{1/5}\Lambda_H=2.29\,,
 \label{NJL para}
\end{align}
which are satisfied for the SM hadrons; hence we assume that these relations remain unchanged for $\Lambda_H\gg 200$ MeV. 

\subsection{Mass spectrum}\label{Mass spectrum}
\label{subsec:mass_spec}
\noindent
\begin{figure}[h]
\begin{center}
\includegraphics[width=0.7\textwidth]{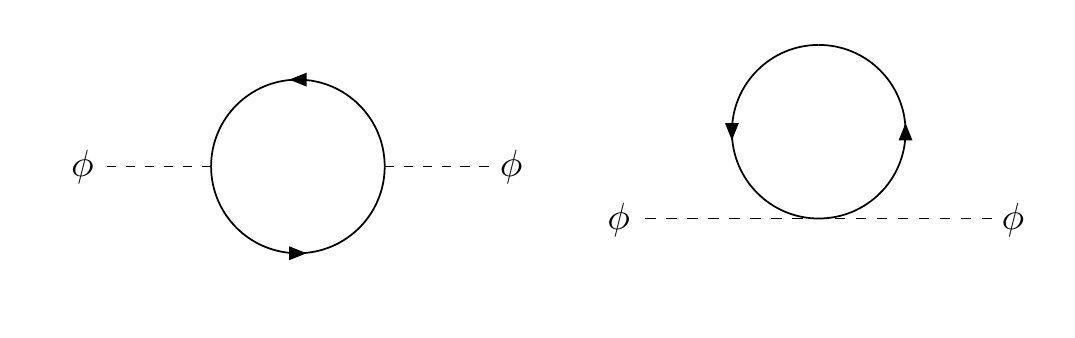}
\caption{One-loop diagrams for
$\Gamma_{\mathrm{DM}}$.}
\label{fig4}
\end{center}
\end{figure}
Once the VEVs 
 $\langle \sigma \rangle$ and $\langle S \rangle$
 (before the EW breaking) are obtained, 
the scalar mass spectrum can be calculated from 
the corresponding two point functions at one-loop
with the hidden fermions.
The VEVs
can be computed from
\be
V_\text{S+NJL}(S,\sigma) &=&\frac{1}{4} \lambda_S\,S^4+V_\text{NJL}\,,
\label{VSNJL}
\ee
where $V_\text{NJL}$ is given in (\ref{eq:Vnjl}).
The CP even scalars $\sigma$ and  $S$  mix with each other;
we will, however, neglect the mixing and use the tree-level mass
for $S$:
\be
m_S^2 &=& 3 \lambda_S v_S^2~~~(v_S=\langle S \rangle)\,.
\label{mS}
\ee
To obtain the mass of DM candidate(s) we have to compute one-loop diagrams
of  \cref{fig4}
to obtain \cite{Holthausen:2013ota}
\begin{align}
\label{GammaDM}
\Gamma_{\mathrm{DM}}(p^2)&=-\frac{1}{2G}+\frac{G_D\left<\sigma\right>}{8G^3}+\left(1-\frac{G_D\left<\sigma\right>}{8G^2}\right)^2 2n_cI_{\phi^2}(p^2,\left<M\right>,\Lambda_{\mathrm{H}}) +\frac{G_D}{G^2}n_cI_V(\left<M\right>,\Lambda_{\mathrm{H}})\,,
\end{align}
where
\al{\left<M\right>
= \langle \sigma \rangle
+y \langle S \rangle-\frac{G_D}{8G^2}\langle \sigma \rangle^2\,,
\label{VEVM}}
and
 the loop functions are given by
\begin{align}
  I_{V}(m,\Lambda) & = \int_{\Lambda}  \frac{d^4 k}{i(2\pi)^4}\frac{m}{(k^2-m^2)}=-\frac{1}{16\pi^2}m\left[ \Lambda^2-m^2\ln \left( 1+\frac{\Lambda^2}{m^2}\right) \right]\,,\\
  I_{\phi^2}(p^2,m,\Lambda) &= \int_{\Lambda}  \frac{d^4 k}{i(2\pi)^4}\frac{\mathrm{Tr}(\Slash{k}-\Slash{p}+m)\gamma_5(\Slash{k}+m)\gamma_5}{((k-p)^2-m^2)(k^2-m^2)}\nn\\
 & =\frac{1}{4 \pi^2}\left[ \Lambda^2+(p^2/2-m^2)\ln (1+\Lambda^2/m^2)
-\sqrt{(4 m^2-p^2)p^2} \arctan\frac{1}{\sqrt{(4 m^2/p^2-1)}}\right.\nn\\
& \left.+\frac{(2 \Lambda^2 +4 m^2-p^2)}{\sqrt{(4\Lambda^2/p^2+4 m^2/p^2-1)}}
\arctan\frac{1}{\sqrt{(4\Lambda^2/p^2+4 m^2/p^2-1)}}
\right]\,.
 \end{align}
Then we can calculate the DM mass from  
\be
\Gamma_{\mathrm{DM}}(m_{\mathrm{DM}}^2)=0\,,
\label{mDM}
\ee
and the wave function renormalization constant $Z$ from
\be
Z^{-1} &=&\left.\frac{d \Gamma_{\mathrm{DM}}(p^2)}{d p^2}
\right|_{p^2=m_{\mathrm{DM}}^2}\,.
\label{Z}
\ee

\begin{figure}[h]
\begin{center}
\includegraphics[width=0.4\textwidth]{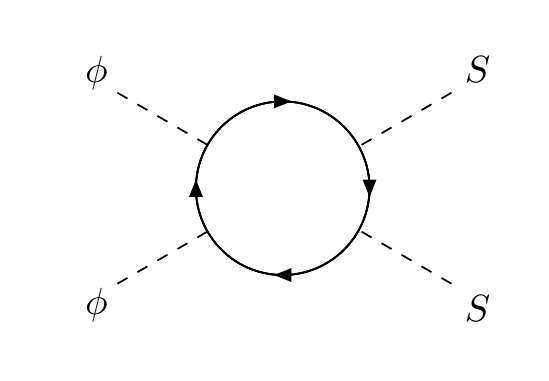}
\includegraphics[width=0.4\textwidth]{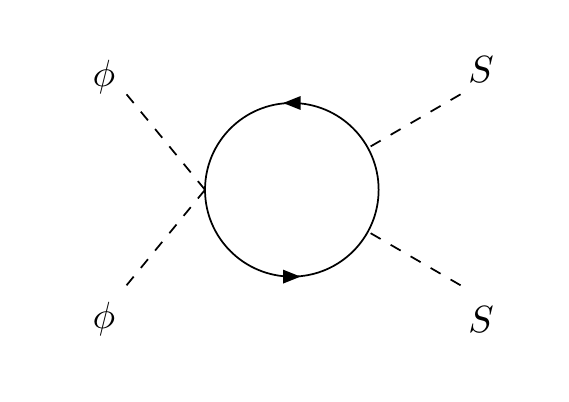}
\caption{One-loop diagrams for $G_{\phi^2S^2}$.}
\label{fig5}
\end{center}
\end{figure}

\begin{figure}[h]
\begin{center}

\includegraphics[width=0.4\textwidth]{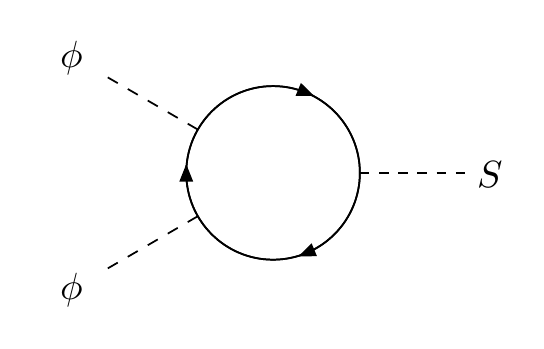}
\includegraphics[width=0.4\textwidth]{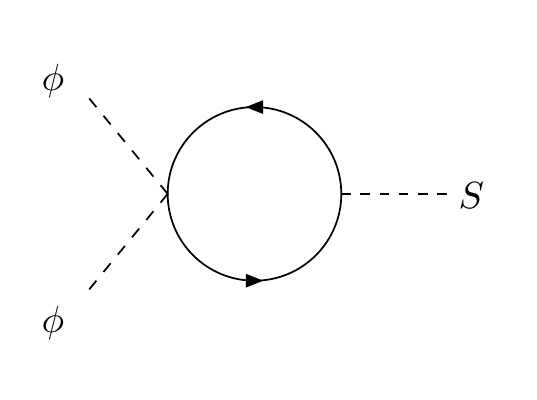}

\caption{One-loop diagrams for $G_{\phi^2S}$.}
\label{fig6}
\end{center}
\end{figure}

\subsection{Dark matter coupling with $S$}
\label{subsec:DM-couple}
\noindent
The diagrams in \cref{fig5,fig6} show
dark matter interactions with the singlet $S$.
We set the external momenta to zero in order to obtain 
their local form (see Ref.\,\cite{Ametani:2015jla}):
\be
{\cal L}_{\phi S} &=& 
\frac{1}{4}G_{\phi^2 S^2}~  \phi_a \phi_a S^2+
\frac{1}{2}G_{\phi^2 S}~  \phi_a \phi_a S\,,
\label{LphiS}
\ee
where
\be
 G_{\phi^2 S^2} &=&
-  Z \,n_c\left(1-\frac{G_D}{8G^2}
\langle\sigma\rangle\right)^2 y^2 
\left[
2 I^{2A}_{\phi^2 S^2}(\left<M\right>,\Lambda_H)+
 I^{2B}_{\phi^2 S^2}(\left<M\right>,\Lambda_H)  \right]\nn\\
& & -  \frac{1}{2}Z\,n_c\left(\frac{G_D}{8G^2}\right)
y^2\, I^{C}_{\phi^2 S^2}(\left<M\right>,\Lambda_H)~,
\label{Gp2S2}\\
G_{\phi^2 S} &=&
-2  Z\,n_c\left(1-\frac{G_D}{8G^2}
\langle\sigma\rangle\right)^2y \,
I^{2A}_{\phi^2 S}(\left<M\right>,\Lambda_H)-
2  Z\, n_c\left(\frac{G_D}{8G^2}\right)y\,I^{B}_{\phi^2 S}(\left<M\right>,\Lambda_H)~\label{Gp2S}\,,
\ee
and $\left<M\right>$ and $Z$ are given in (\ref{VEVM}) and  (\ref{Z}), respectively.
The loop functions are given by
\al{
 I^{2A}_{\phi^2 S^2}(m,\Lambda) &=\left.
 (-1)\int_{\Lambda} \frac{d^4 l}{i(2\pi)^4}
\frac{\mbox{Tr}
(\slashed{l}+m)\gamma_5
(\slashed{l}-\slashed{p}'+m)
(\slashed{l}+\slashed{p}-\slashed{k}+m)
(\slashed{l}+\slashed{p}+m)\gamma_5}{(l^2-m^2)((l-p')^2-m^2)
((l+p-k)^2-m^2)((l+p)^2-m^2)}\right|_{p=p'=k=0}\nn\\
& +(k\leftrightarrow k')
=-\frac{1}{2\pi^2}\left[
2-\ln (\Lambda^2/m^2)
+O(m^2/\Lambda^2)\right]\,, \label{loop1}\\
 I^{2B}_{\phi^2 S^2}(m,\Lambda) &=\left.
  (-1)\int_{\Lambda} \frac{d^4 l}{i(2\pi)^4}
\frac{\mbox{Tr}
(\slashed{l}+m)
(\slashed{l}+\slashed{k}'+m)\gamma_5
(\slashed{l}+\slashed{p}-\slashed{k}+m)
(\slashed{l}+\slashed{p}+m)\gamma_5}{(l^2-m^2)((l+k')^2-m^2)
((l+p-k)^2-m^2)((l+p)^2-m^2)}\right|_{p=k=k'=0}~\nn\\
& +(k\leftrightarrow k')~
=-\frac{1}{2\pi^2}\left[
-1+\ln (\Lambda^2/m^2)
+O(m^2/\Lambda^2)\right]\,,~\\
   I^{C}_{\phi^2 S^2}(m,\Lambda) &=\left.
   (-1)\int_{\Lambda} \frac{d^4 l}{i(2\pi)^4}
\frac{\mbox{Tr}
(\slashed{l}-\slashed{k}'+m)(\slashed{l}+m)
(\slashed{l}+\slashed{k}+m)}{((l+k)^2-m^2)
(l^2-m^2)((l-k')^2-m^2)}\right|_{k'=k=0}~
+(k\leftrightarrow k')\nn\\
& = \frac{m}{2\pi^2}\left[
5-3\ln (\Lambda^2/m^2)
+O(m^2/\Lambda^2)\right]\,,\\
      I^{2A}_{\phi^2 S}(m,\Lambda) &=\left.
      (-1)\int_{\Lambda} \frac{d^4 l}{i(2\pi)^4}
\frac{\mbox{Tr}
(\slashed{l}+p+m)
\gamma_5(\slashed{l}+m)\gamma_5
(\slashed{l}-\slashed{p}'+m)}{((l+p)^2-m^2)(l^2-m^2)
((l-p')^2-m^2)}\right|_{p=p'=0}~+(p\leftrightarrow p')\nn\\
&=
\frac{m}{4\pi^2}\left[-1+ 
\ln(\Lambda^2/m^2)+O(m^2/\Lambda^2)\right]\,,\\
         I^{B}_{\phi^2 S}(m,\Lambda) &=\left.
         (-1)\int_{\Lambda} \frac{d^4 l}{i(2\pi)^4}
\frac{\mbox{Tr}
(\slashed{l}+\slashed{p}+\slashed{p}'+m)
(\slashed{l}+m)}{((l+p+p')^2-m^2)(l^2-m^2)}
\right|_{p=p'=0}\nn\\
&= \frac{1}{4\pi^2}
\left[\Lambda^2+2 m^2-3 m^2\ln(\Lambda^2/m^2)+
m^2 \,O(m^2/\Lambda^2)\right]\,.\label{loop5}
}
\section{Dark Matter and Freeze-in}
\label{sec:freeze-in-DM}
\begin{figure}
  \centering
  \begin{tabular}{cc}
    \includegraphics[width=0.48\textwidth]{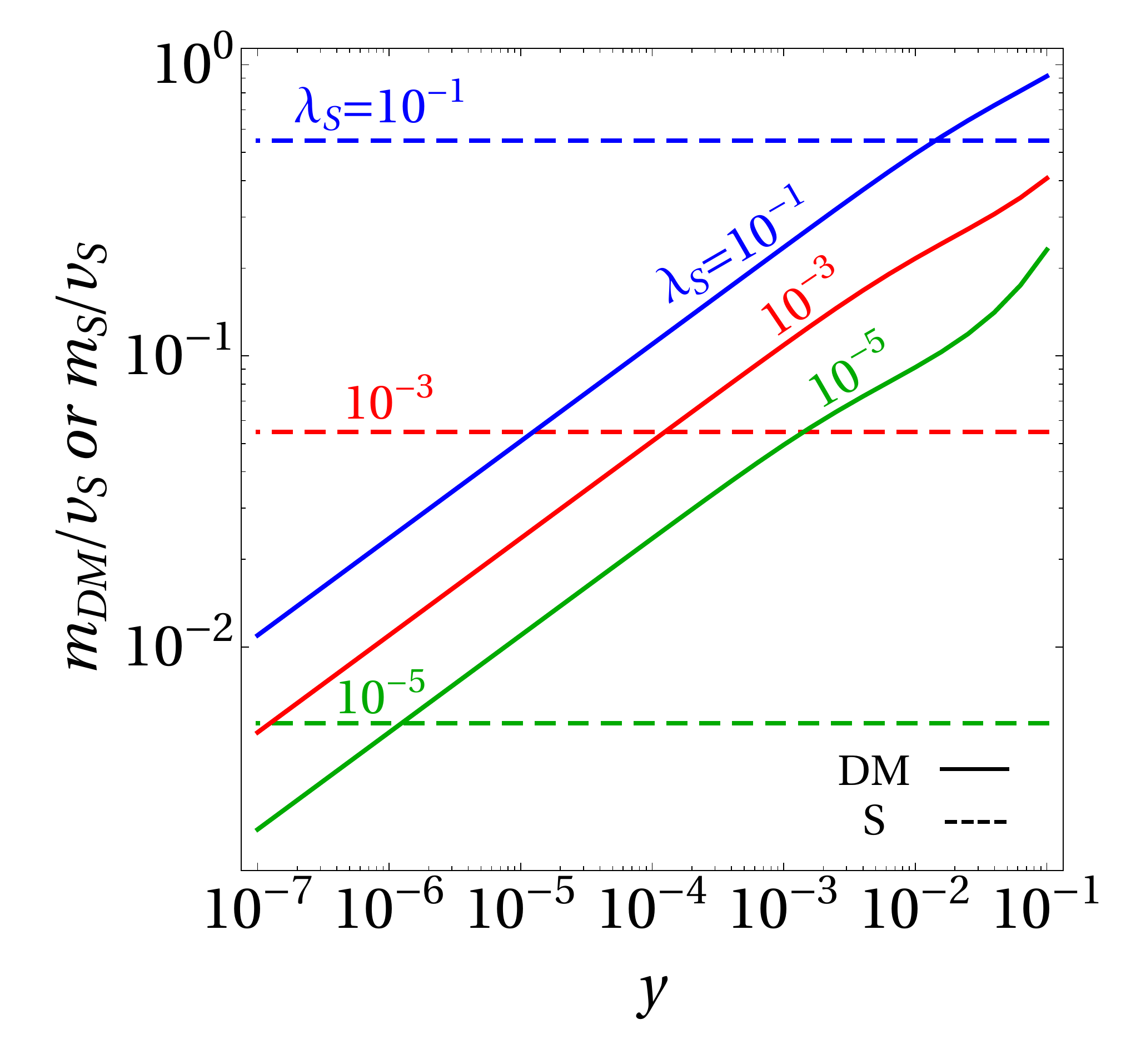} &
    \includegraphics[width=0.48\textwidth]{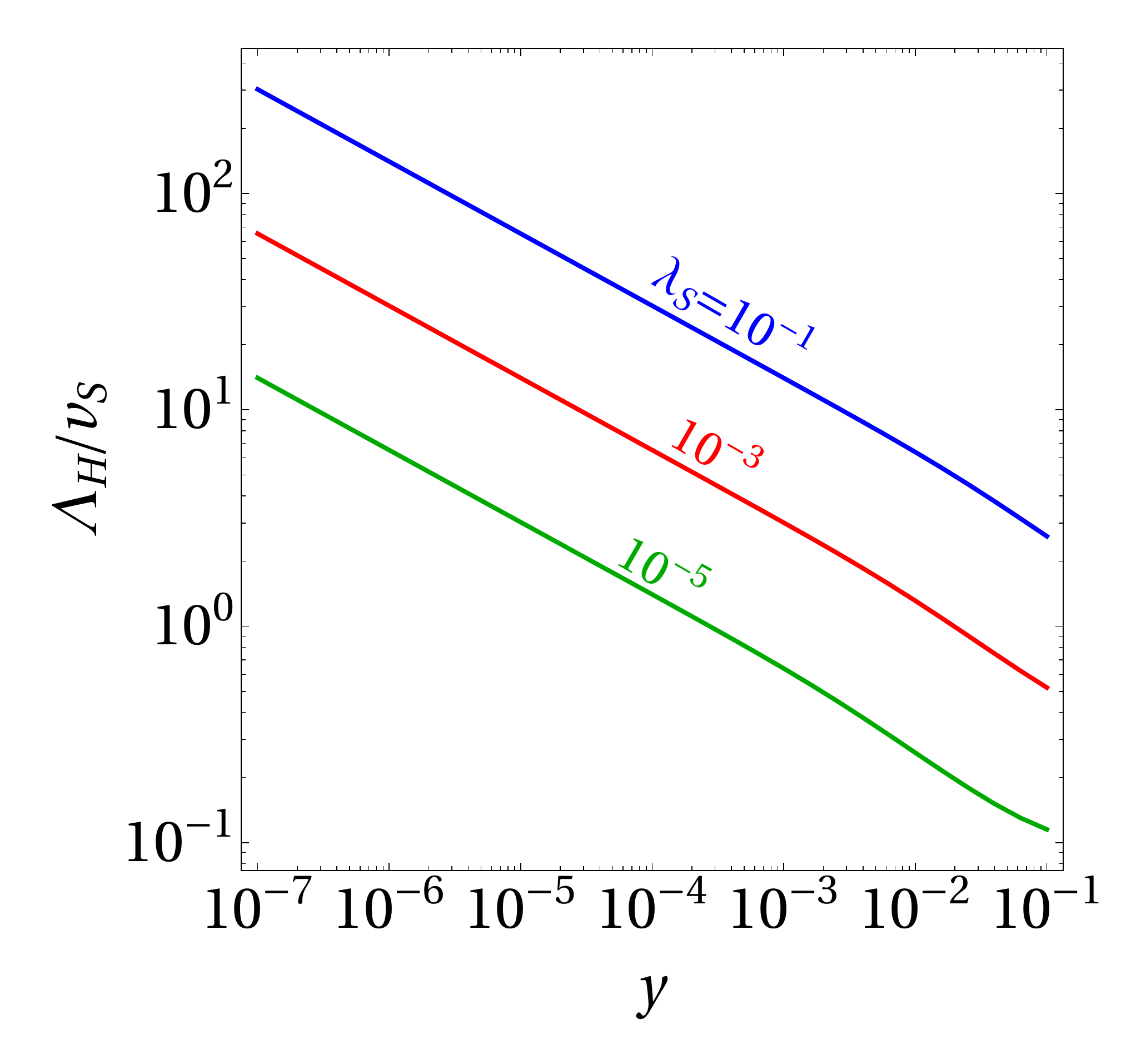} \\
  \end{tabular}
\caption{\emph{Left}:\,$m_\text{DM}$ and $m_S$ shown against the Yukawa coupling $y$,
where the masses are normalized by $m_N/y_M=v_S$. 
The solid and dashed lines stand for $m_\text{DM}/(m_N/y_M)$
and $m_S/(m_N/y_M)$, respectively.
\emph{Right}:\,$\Lambda_H/v_S$ against the Yukawa coupling $y$.
In both panels we show curves corresponding to $\lambda_S=10^{-1},\,10^{-3},\,\text{and}\, 10^{-5}$\,.
}
\label{mDM-mS}
\end{figure}

\noindent
 The quasi NG bosons in the hidden sector are our DM candidates primarily due to their stability  stemming from the unbroken $SU(3)_V$ flavor symmetry. However, they are very heavy as shown in the left panel of Fig.\,\ref{mDM-mS}; there the masses of $m_\text{DM}$ (solid lines) and $m_S$ (dashed) are shown with respect to the absolute scale of the hidden sector that is fixed by 
 $v_S=m_N/y_M=5\times 10^7/y_M$ GeV.
As we see from this panel, $m_\text{DM}$ is
larger than $2.5\times 10^6$ GeV for $y \gsim 10^{-6},
 \lambda_S \gsim 10^{-5}$ and $y_M \lsim 0.1$. In our scans we will focus on the mass spectrum $m_{\text{DM}}\simeq m_S > m_N$.
 
Clearly, such NG bosons are too heavy for a conventional freeze-out at measured value of DM number density \cite{Griest:1989wd} (for freeze-out production in the presence of hidden strongly interacting sector see \cite{Dondi:2019olm,Ahmed:2020hiw}). 
 The situation at hand is actually very similar
 to that considered in Refs.\,\cite{Berlin:2016vnh,Berlin:2016gtr}: $X$ ($Y$) 
 in Ref.\,\cite{Berlin:2016gtr} would correspond to our $\phi$ ($S$).  
 However, our $S$ is not ``highly decoupled" from the SM sector; it decays with the lifetime $\tau_S 
 \lsim  4\times 10^{-24}$ s for $y_M, \lambda_S > 10^{-5}$, and this should be contrasted with the lifetime of $Y\sim \mathcal{O}(1)$ s \cite{Berlin:2016vnh,Berlin:2016gtr}.
 The bottom line is that our scenario is not compatible with neither conventional freeze-out nor its derivatives. Hence, we 
will therefore explore the possibility of the freeze-in mechanism \cite{Hall:2009bx}. Since our dark sector is at masses $\gtrsim 10^7$ GeV, the temperatures where the freeze-in 
is most efficient will be of similar order.
 
 An important question to ask is whether the hidden sector is  thermalized. If so, then it undergoes the chiral phase transition at 
 the critical temperature $T_C$, which  is 
 about $10^{-1}\times \Lambda_H$ \cite{Holthausen:2013ota, Aoki:2019mlt}. In passing, we mention that $T_C \sim 0.1$ GeV  in QCD with nearly massless quarks
 \cite{Pisarski:1983ms, DeTar:2009ef, Meyer:2015wax,Jin:2017jjp,Helmboldt:2019pan}. 
 In the right panel of \cref{mDM-mS} we show $\Lambda_H$ in  the same parameter space as for the left panel discussed above.
 We see that, in contrast to $m_\text{DM}$, the smaller $y$ is, the larger $\Lambda_H$ becomes; it
 is larger than $\sim 5\times 10^7$ GeV
 for $y \lsim 0.1, \lambda_S \gsim 10^{-5}$ and $y_M \lsim 0.1$. So, the critical temperature $T_C$ would be higher than $\sim 5\times 10^6$ GeV. If dark sector temperature 
 reaches $T_C$ during the evolution of the Universe, the hidden sector undergoes a phase transition and NB Goldstone appear in the theory. Notice that at $T>T_C$ such particles are not even present in the model.

If, on the other hand,  the hidden sector is not thermalized 
 above $T_C$, there will be no chiral phase transition;
 the chiral symmetry  is dynamically  broken  
 all the way to present, and the hidden sector is in  confining phase all the time. In this case our DM candidate is present from the beginning.
 Its number density would be zero  if the hidden sector was completely disconnected with the SM, 
 because the SM sector could not heat up the hidden sector
 in such situation\footnote{Assuming that there exits no other
 mechanism that can heat up the hidden sector. This should be contrasted to the assumption of Ref.\,\cite{Berlin:2016gtr} 
 that the SM  and hidden sectors are both thermally populated during post-inflation reheating and they maintain separate temperatures.}.
We recall that there is no direct contact between the hidden sector and the SM sector;
they are only connected  indirectly via mediator $S$, which
has  contact with the SM sector
through the Higgs portal interaction  
$\lambda_{HS}$, which we previously argued to be very tiny.

An indirect contact of $S$ with the SM sector is established by
the Yukawa interaction $y_M$,
while
the  right-handed neutrino $N_R$ is in  direct contact  with the SM thermal bath 
due to the Yukawa interaction $y_\nu$
which is fixed to $\mathcal{O}(10^{-4})$ in order to reproduce light neutrino masses.
Therefore, the connection of the hidden sector  to the SM sector is considerably suppressed. In the absence of inflation-scale effects let us further mention that the thermalization of $S$ would yield the thermalization of the hidden sector. We have however found that 
this is not satisfied for considered range of parameters as will be shown in \cref{subsec:results}. 
Hence, in what follows we proceed with the assumption that the hidden sector is cold and in particular not reheated during inflation.

\subsection{Boltzmann equations and thermally averaged cross sections}
\label{subsec:boltzmann}
\noindent
The relevant physical degrees of freedom to be considered 
are $\phi, S, N_R$ as well as the SM fields which are in thermal equilibrium.
The set of Boltzmann equations to be solved reads
  \be
& &\frac{d Y_\phi}{dx}=
-\left[\left(\frac{\pi g_*}{45} \right)^{1/2}\frac{\mu M_{\rm PL}}{x^2} \right]
\left\{<\!\sigma (\phi\phi;H^\dag H) v\!>
\left(  Y_\phi Y _\phi-\bar{Y}_{\phi}\bar{Y}_{\phi}\right)+\right.\nn\\
& &\left.
<\!\sigma (\phi \phi ;SS)v\!>\!\!\,\left(  Y_\phi  Y _\phi -\frac{Y_S Y_S}
{\bar{Y}_{S}\bar{Y}_{S}} \bar{Y}_{\phi }\bar{Y}_{\phi }
\right)
+<\!\sigma (\phi \phi ;N_R N_R)v\!>\!\!\left(  Y_\phi  Y _\phi 
-\frac{Y_{N_R} Y_{N_R}}
{\bar{Y}_{N_R}\bar{Y}_{N_R}} \bar{Y}_{\phi }\bar{Y}_{\phi }
\right)\right\}\,,\\
& &\frac{d Y_S}{dx}=
- \left[\left(\frac{\pi g_*}{45} \right)^{1/2}\frac{\mu M_{\rm PL}}{x^2} \right]
\left\{<\!\sigma (SS;H^\dag H) v\!>
\left(  Y_S Y _S-\bar{Y}_{S}\bar{Y}_{S}\right)+\right.\nn\\
& &\left.
 <\!\sigma (SS;N_R N_R)v\!>\!\!\left(  Y_S Y _S-\frac{Y_{N_R} Y_{N_R}}
{\bar{Y}_{N_R}\bar{Y}_{N_R}} \bar{Y}_{S}\bar{Y}_{S}\right)
-8 <\!\sigma (\phi \phi ;SS)v\!>\!\!\left(  Y_\phi  Y _\phi -\frac{Y_S Y_S}
{\bar{Y}_{S}\bar{Y}_{S}} \bar{Y}_{\phi }\bar{Y}_{\phi }
\right)\right\}
\nn\\
& &- \left[\left(\frac{90}{8\pi^3 g_*}
\right)^{1/2}\frac{x M_{\rm PL}}{\mu^2} \right]
\left\{ <\!\Gamma (S;N_R N_R)\!>\!\!\left(  
Y_S -\frac{Y_{N_R}Y_{N_R}}
{\bar{Y}_{N_R}\bar{Y}_{N_R}}\bar{Y}_{S}
\right)+<\!\Gamma (S;H^\dag H)\!>\!\!\left(  
Y_S -\bar{Y}_{S}
\right)\right\}~,\\
& &\frac{d Y_{N_R}}{dx}=
-\left[\left(\frac{\pi g_*}{45} \right)^{1/2}\frac{\mu M_{\rm PL}}{x^2} \right]
\left\{\left(<\!\sigma (N_R N_R;\tilde{H}^\dag \tilde{H}) v\!>+
<\!\sigma (N_R N_R;\bar{L} L) v\!>\right) \right. \nn \\ & &\left.
\left(  Y_{N_R} Y _{N_R}-\bar{Y}_{N_R}\bar{Y}_{N_R}\right)
-(8/3) <\!\sigma (\phi \phi ;N_R N_R)v\!>\!\!\left(  Y_\phi  Y _\phi 
-\frac{Y_{N_R} Y_{N_R}}
{\bar{Y}_{N_R}\bar{Y}_{N_R}} \bar{Y}_{\phi }\bar{Y}_{\phi }
\right)
-(1/3)<\!\sigma (SS;N_R N_R)v\!>\!\!
\right. \nn \\ & &\left.
\left(  Y_S Y _S-\frac{Y_{N_R} Y_{N_R}}
{\bar{Y}_{N_R}\bar{Y}_{N_R}} \bar{Y}_{S}\bar{Y}_{S}\right)\right\}-\left[\left(\frac{90}{8\pi^3 g_*}
\right)^{1/2}\frac{x M_{\rm PL}}{\mu^2} \right]
\left\{-(1/3)<\!\Gamma (S;N_R N_R)\!>\!\!\left(  
Y_S-\frac{Y_{N_R}Y_{N_R}}
{\bar{Y}_{N_R}\bar{Y}_{N_R}}\bar{Y}_{S}
\right)+\right. \nn \\ & &\left.<\!\Gamma (N_R;L H)\!>\!\!\,\,\left(  Y_{N_R} -\bar{Y}_{N_R}
\right) \right\}~,
\label{boltz3}
\ee
where $M_{\rm PL}$ is non-reduced Planck mass, $g_*$ is the number of degrees of freedom fixed to $106.75$, $1/\mu=1/m_\text{DM}+1/m_S+1/m_N$, $x=\mu/T$,
and we assume the spectrum $m_\text{DM} > m_S > m_N$.
Here $Y_\phi$ is the number density of 
one $\phi$ divided by entropy, dubbed yield in what follows ($Y_{\phi}=Y_{\phi_1}=\cdots=Y_{\phi_8}$
because of the unbroken $SU(3)_V$). 
Similarly, $Y_{N_R}$ is the yield of 
one $N_R$, we also have
$Y_{N_R}=Y_{N_{R1}}=Y_{N_{R2}}=Y_{N_{R3}}$.
Note, however, that the internal degrees of freedom 
for $N_R$ are counted as two. The yields denoted with the bar 
represent values in equilibrium. When solving the Boltzmann equations, we apply the initial condition where all yields are equal to zero at starting temperature that exceeds DM mass.

\begin{figure}[h]
\begin{center}

\includegraphics[width=0.8\textwidth]{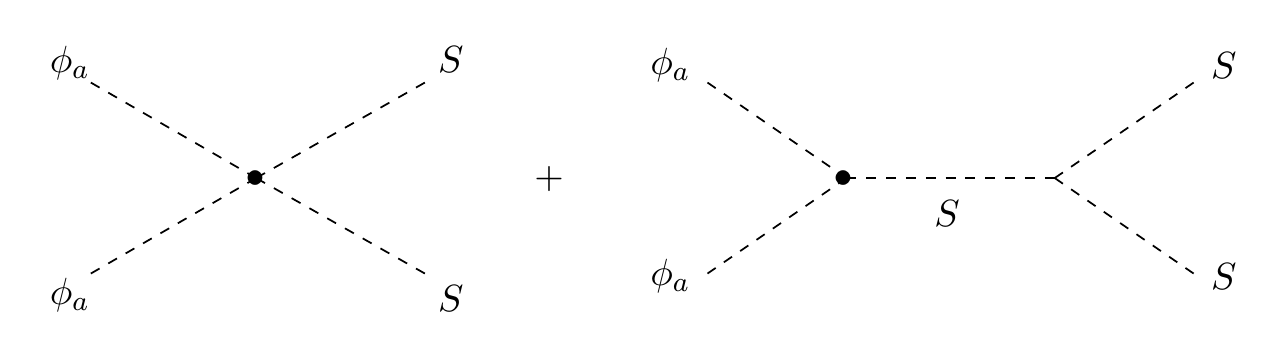}
\includegraphics[width=0.4\textwidth]{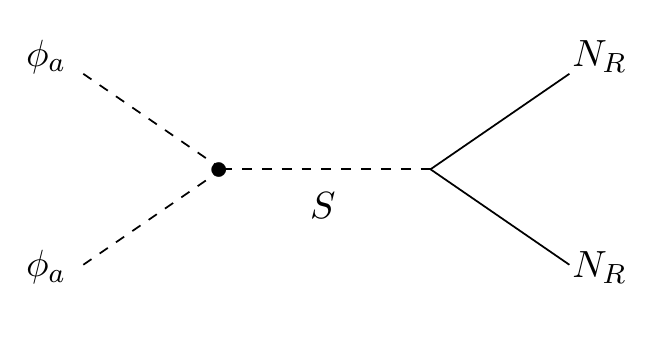}
\includegraphics[width=0.4\textwidth]{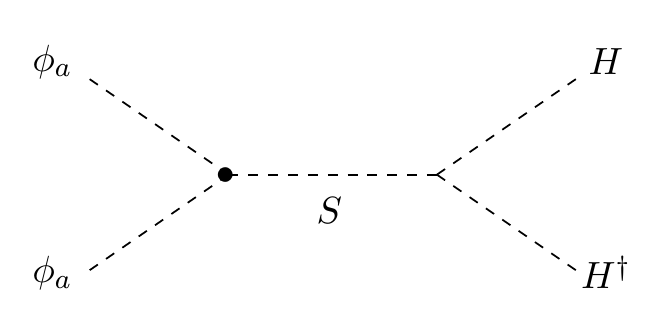}

\caption{Annihilation diagrams for $\phi$, where  $\bullet$ indicates
the   effective coupling (\ref{Gp2S2}) for the upper-left 
diagram  and (\ref{Gp2S}) for the other diagrams.
They contribute to $\sigma (\phi\phi;SS)~\mbox{(upper)},
\,\sigma (\phi\phi;N_RN_R)$ 
(lower left) and $ \sigma (\phi\phi;H^\dag H)$ (lower right) , respectively.
}
\label{dmdm}
\end{center}
\end{figure}

\begin{figure}[h]
\begin{center}
\includegraphics[width=0.6\textwidth]{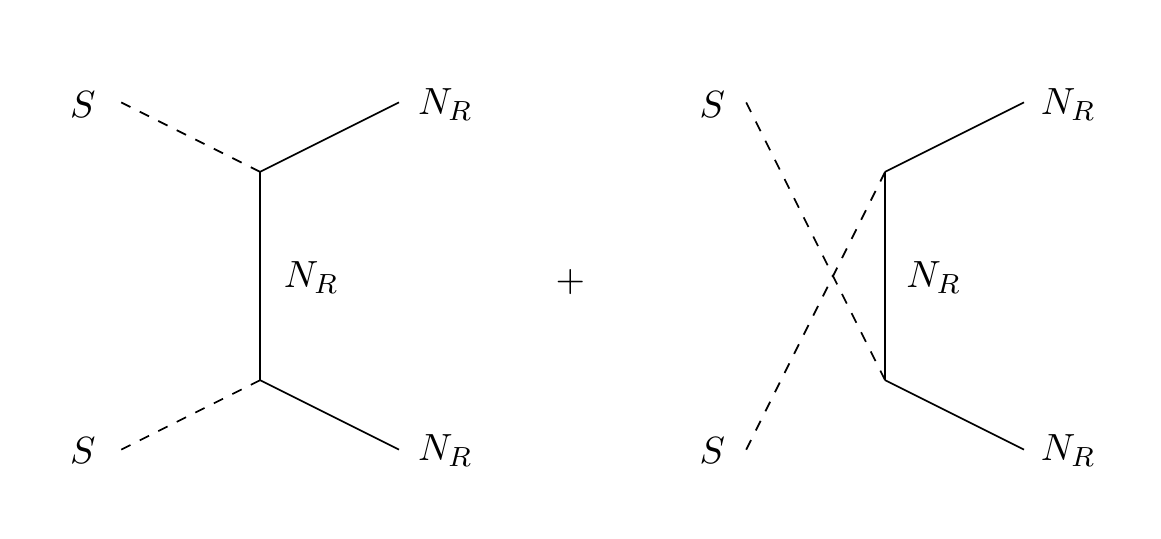}
\includegraphics[width=0.37\textwidth]{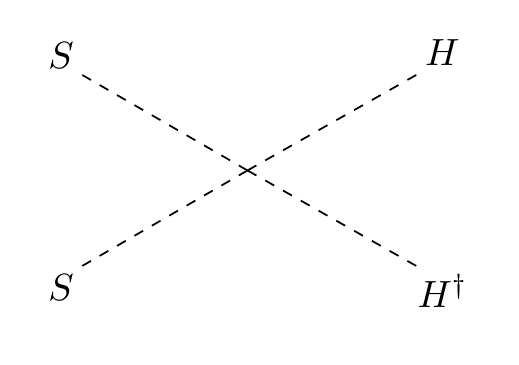}
\caption{Annihilation diagrams for $S$, which contribute to
$\sigma (SS;N_R N_R)$ (left, middle) and $ \sigma (SS;H^\dag H)$
(right), respectively.}
\label{SS}
\end{center}
\end{figure}

\begin{figure}[h]
\begin{center}
\includegraphics[width=0.7\textwidth]{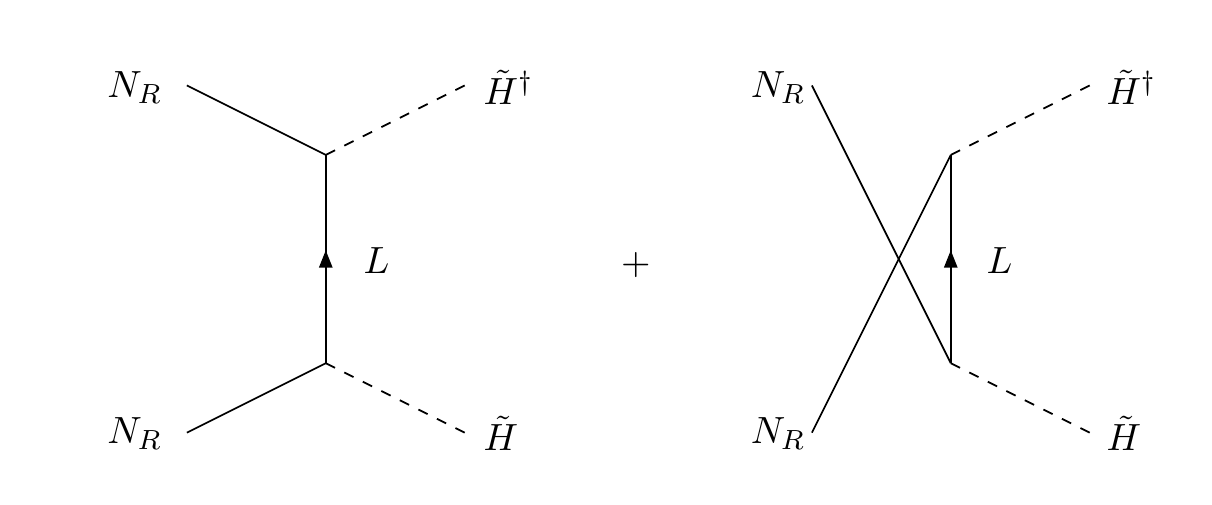}

\includegraphics[width=0.7\textwidth]{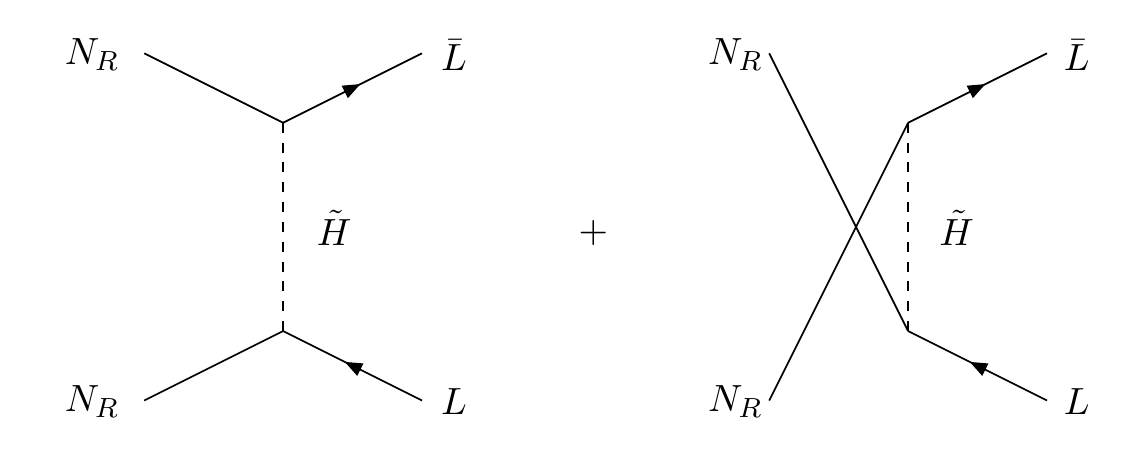}
\caption{Annihilation diagrams for $N_R$, which contribute to
$\sigma (N_R N_R;\tilde{H}^\dag \tilde{H})$ 
(upper) and $ \sigma (N_RN_R;\bar{L}L)$ (lower), respectively.}
\label{NN}

\end{center}
\end{figure}

\begin{figure}[h]
\begin{center}
\includegraphics[width=0.4\textwidth]{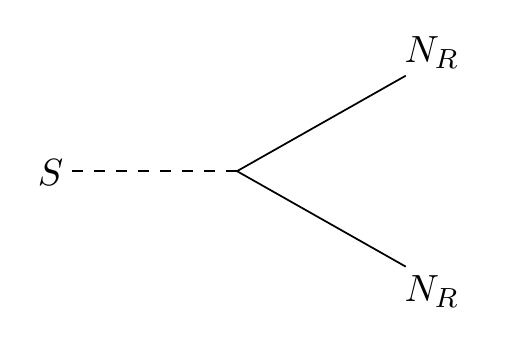}
\includegraphics[width=0.4\textwidth]{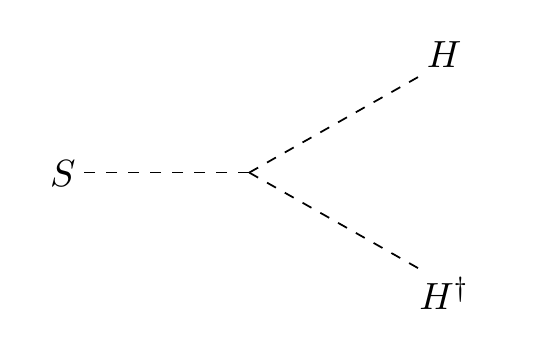}
\caption{Diagrams for the $S$ decay into
$N_R N_R$ (left) and $H^\dag H$ (right).}
\label{S-decay}
\end{center}
\end{figure}

\begin{figure}[h]
\begin{center}
\includegraphics[width=0.8\textwidth]{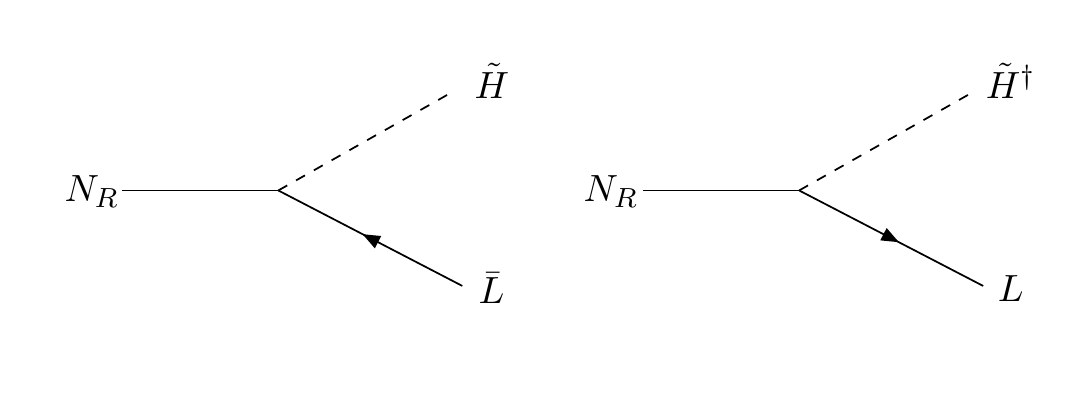}
\caption{Diagrams for the $N_R$ decay into
$\tilde{H} \bar{L}$ (left) and $\tilde{H}^\dag  L$ (right).}
\label{N-decay}
\end{center}
\end{figure}

The Feynman diagrams for all relevant terms in (\ref{boltz3}) are shown in \cref{dmdm,SS,NN,S-decay,N-decay}.


The thermally averaged cross section is defined as \cite{Gondolo:1990dk}
\al{
\langle \sigma v \rangle
&=\frac{\int d^3 p_1 d^3p_2
\, (\sigma v)\, \exp(-E_1/T)\exp(-E_2/T)}{
\int d^3 p_1d^3 p_2
\exp(-E_1/T)\exp(-E_2/T)}
\,\nn\\
&=\frac{\int d\Omega_1d\Omega_2dE_1dE_2  p_1 p_2
\,\left[E_1 E_2(\sigma v)\right]\, \exp(-E_1/T)\exp(-E_2/T)}{
\left(4\pi\int d E_1 p_1 E_1
\exp(-E_1/T)\right)^2}\,,
\label{sigma-v}}
while the thermally averaged decay rate reads
\al{
\langle \Gamma\rangle
&=\frac{\int d^3 p
\,\Gamma\,\exp(-E/T)}{\int d^3 p\exp(-E/T)}\,=\frac{\int dE\, p E
\,\Gamma\,\exp(-E/T)}{\int dE  \,p E \exp(-E/T)}\,.
\label{Gamma}
}
Here, $E_1$ and $E_2$ are the energies of the annihilating
particles, and $E$ is the energy of the decaying particle.
Note that  $E_1 E_2(\sigma v)$ can be expressed in a covariant form as \cite{Gondolo:1990dk}
\al{
E_1 E_2(\sigma v)&=\sigma F(m),~\mbox{where}~
F(m)=\sigma \sqrt{s (s-4m^2)}/2\,,
}
and $s$ is the Mandelstam variable in units of energy squared.
Since $\sigma F(m)$
is a function of $s$, dubbed $\tilde{\sigma}(s)$, 
the integration (\ref{sigma-v})
can not be carried out analytically in general.
However, if $\tilde{\sigma}(s)$ is a constant independent of $s$, for instance,
we can perform the integration
analytically. So, to proceed, we approximate the integral 
by replacing  $s$ by $4 m_\text{DM}^2$ in $\tilde{\sigma}(s)$ for the DM annihilation
and also by $4 m_S^2$
for the $S$ annihilation.
In this way the high temperature behavior of the thermally averaged cross
section, $\langle \sigma v\rangle\propto 1/T^2$, can be partly 
included in the s-wave contribution. This procedure greatly simplifies numerical evaluation of (\ref{boltz3}); note that we have checked its robustness by making an exact evaluation of (\ref{sigma-v}) for one of the processes and found minor differences at temperatures where the DM production occurs.
As argued above, this simplification now allows us to perform the thermal averaging analytically, and
here are the computed expressions\footnote{We neglect below 
$\sigma (N_R N_R;\tilde{H}^\dag\tilde{H})$
and $\sigma (N_R N_R;\bar{L}L)$; their contribution
is proportional to $y_\nu^4=10^{-14}$, while
$\Gamma (N_R;LH)\propto y_\nu^2=10^{-7}$.
}:
\al{
\langle \sigma (\phi\phi;SS) v \rangle&\simeq
\frac{\left[G_{\phi^2S^2}+6 \lambda_S v_S
 G_{\phi^2S}\Delta(4m_\text{DM}^2)\right]^2}{64 
 \pi m_\text{DM}^2}
\left(1-\frac{m_S^2}{m_\text{DM}^2}\right)^{1/2}\times \left[\frac{K_1(m_\text{DM}/T)}{K_2(m_\text{DM}/T)}\right]^2\,,
\label{phiphiSS1}\\
\langle \sigma (\phi\phi;H^\dag H) v \rangle&\simeq
 \frac{ \left[v_S \lambda_{HS}G_{\phi^2S}
 \Delta(4m_\text{DM}^2)\right]^2}{16\pi 
 m_\text{DM}^2}\left[\frac{K_1(m_\text{DM}/T)}
 {K_2(m_\text{DM}/T)}\right]^2\,,\\
\langle \sigma (\phi\phi;N_R N_R) v \rangle&\simeq
3\times \frac{[G_{\phi^2S}\, y_M
\Delta(4m_\text{DM}^2)]^2}{8\pi }
\left(1- \frac{m_N^2}{m_\text{DM}^2}\right)^{3/2}
\left[\frac{K_1(m_\text{DM}/T)}{K_2(m_\text{DM}/T)}\right]^2\,,
\label{phiphiNN1}\\
\langle \sigma (SS;N_R N_R) v \rangle&\simeq
\frac{3}{2\pi m_S^4}
\left\{(v_S \lambda_S  y_M)^2
-4 v_S \lambda_S   y_M^3 m_N+4 y_M^4 m_N^2\right\}
 \times \nonumber \\ & \left(1-\frac{m_N^2}{m_S^2}\right)^{3/2}\left[\frac{K_1(m_S/T)}{K_2(m_S/T)}\right]^2\,,
\label{SSNN1}\\
\langle \sigma (SS;H^\dag H) v \rangle&
=\frac{\lambda_{HS}^2}{16\pi m_S^2}
\left[\frac{K_1(m_S/T)}{K_2(m_S/T)}\right]^2\,,
\label{SSHH1}\\
\langle \Gamma (S;H^\dag H) \rangle&=
\frac{(v_S \lambda_{HS})^2}{8\pi m_S}
\left[\frac{K_1(m_S/T)}{K_2(m_S/T)}\right]\,,
\label{SHH1}\\
\langle \Gamma (S;N_R N_R) \rangle&=
3\times \frac{y_M^2 m_S}{16\pi}
\left(1-\frac{4m_N^2}{m_S^2}\right)^{3/2}
\left[\frac{K_1(m_S/T)}{K_2(m_S/T)}\right]
\,,
\label{SNN1}\\
\langle \Gamma (N_R;LH) \rangle&=
\frac{m_\nu m_N^2 }{8\pi v_h^2}
\left[\frac{K_1(m_N/T)}{K_2(m_N/T)}\right]\,~(v_h=246\,\mbox{GeV})\,,
\label{NLH1}
}
where $\Delta(4 m_\text{DM}^2)=(4 m_\text{DM}^2-m_S^2)^{-1}$. In calculation we have used
\al{
\int_{m} d E (E^2-m^2)^{1/2}  \exp (-E/T)
&= m T K_1(m/T)\,,\\
\int_{m} d E (E^2-m^2)^{1/2} E \exp (-E/T)
&= m^2 T K_2(m/T)\,.
}
As we see from the above expressions, the approximate
thermally averaged cross section can be obtained from the 
corresponding s-wave cross section by multiplying it
with $\left[K_1(m/T)/K_2(m/T)\right]^2$; the final expression
approaches $1$ as $T$ goes to zero and 
can be approximated as  $m^2/4T^2$ for large $T$. We also note that factors $\left[K_1(m/T)/K_2(m/T)\right]$ appear in thermally averaged decay rate for which made an exact calculation without approximations.

\subsection{Results}
\label{subsec:results}
\begin{figure}[h!]
  \centering
  \begin{tabular}{cc}
    \includegraphics[width=0.7\textwidth]{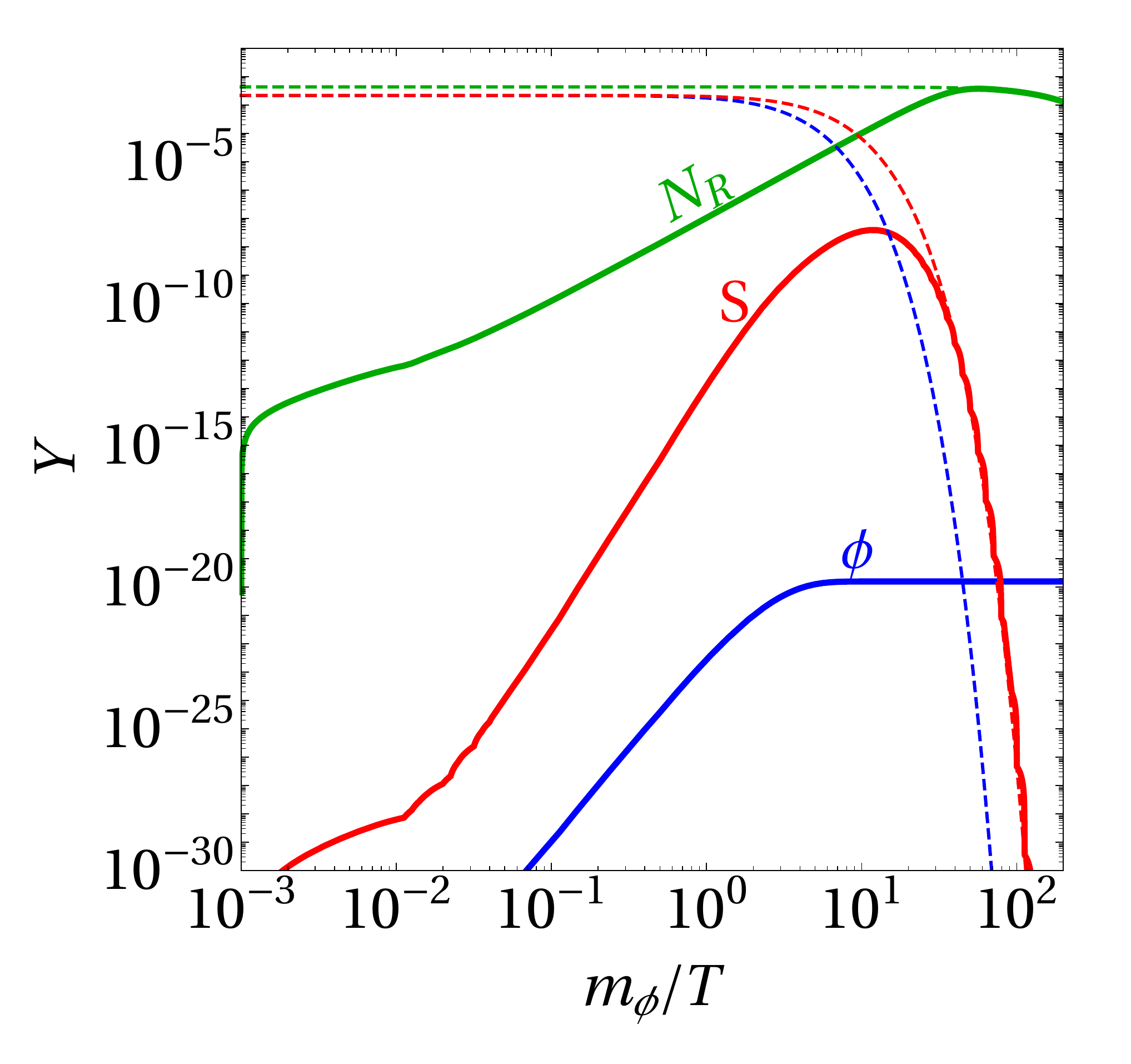}
  \end{tabular}
\caption{Temperature dependence of $\phi$, $N_R$ and $S$ yields
for the benchmark point given in \cref{input}. Corresponding equilibrium values are shown in dashed.}
\label{fig:benchmark}
\end{figure}

\noindent
In this section we will explicitly demonstrate the success of DM production in the model.
Before presenting the results of a scan 
we illustrate our findings by showing a benchmark point for which $\Omega h^2 \simeq 0.12$ and the temperature-dependent yields of all relevant particles involved show a typical behavior. The temperature dependence of $\phi$, $N_R$ and $S$
yields is shown in \cref{fig:benchmark} for a representative 
benchmark point

\al{
m_N &=5\times10^7~\mbox{GeV}\,,m_\nu=0.1~\mbox{eV}\,,
y_\nu =\sqrt{m_\nu m_N}/v_h=2.874\times 10^{-4}\,,\nn\\
y_M & =7\times 10^{-4}\,,\,
y=1.6\times 10^{-4}\,,\,\lambda_S=2.8\times 10^{-4}\,,
\lambda_{HS}=0\,,\nn\\
v_S&=m_N/y_M=(5/7) \times 10^{11}~\mbox{GeV}\,,
\label{input}
}
which gives
\al{
\Lambda_H &=2.60\times10^{11}~\mbox{GeV}\,,
G_{\phi^2S^2}=-1.39\times 10^{-8}
\,,\,G_{\phi^2S}=-4.16\times 10^5~\mbox{GeV}\,,\nn\\
m_\text{DM} & =3.44\times 10^{9}~\mbox{GeV}\,,\,
m_S  =2.07\times 10^{9}~\mbox{GeV}\,,\,
v_\sigma=4.17\times 10^{10}~\mbox{GeV}\,.
}

\begin{figure}[h!]
  \centering
  \begin{tabular}{cc}
    \includegraphics[width=0.33\textwidth]{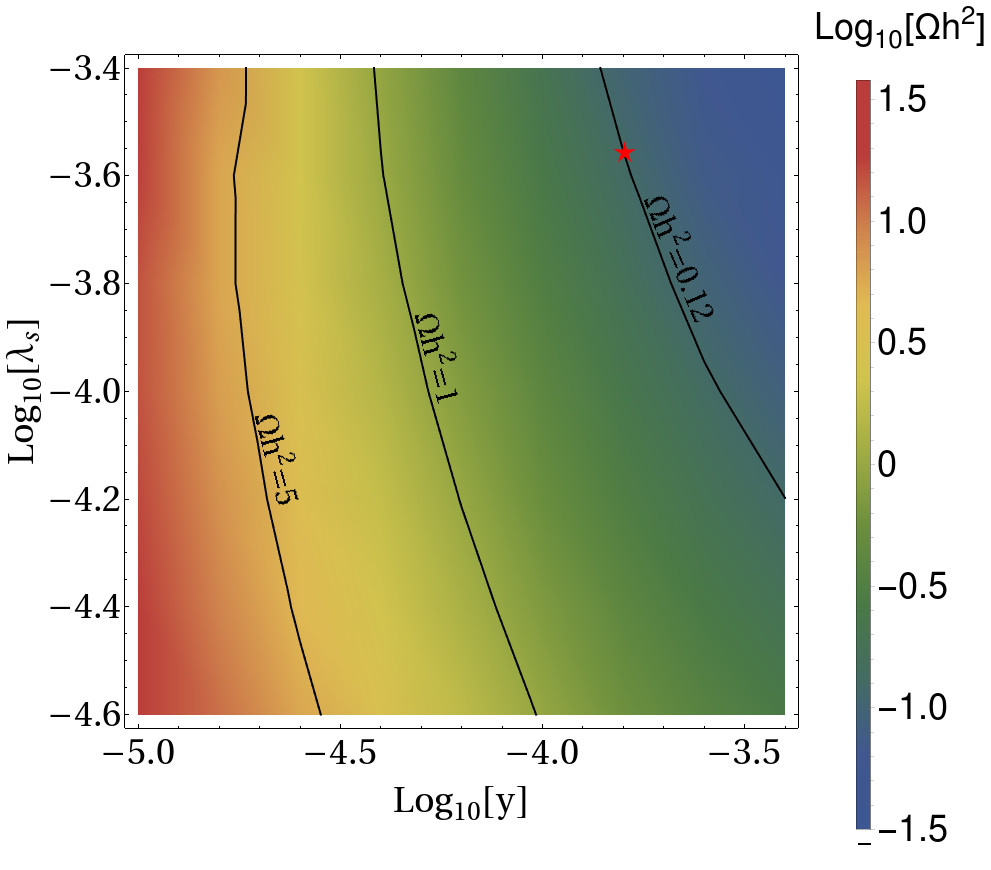} 
   \includegraphics[width=0.33\textwidth]{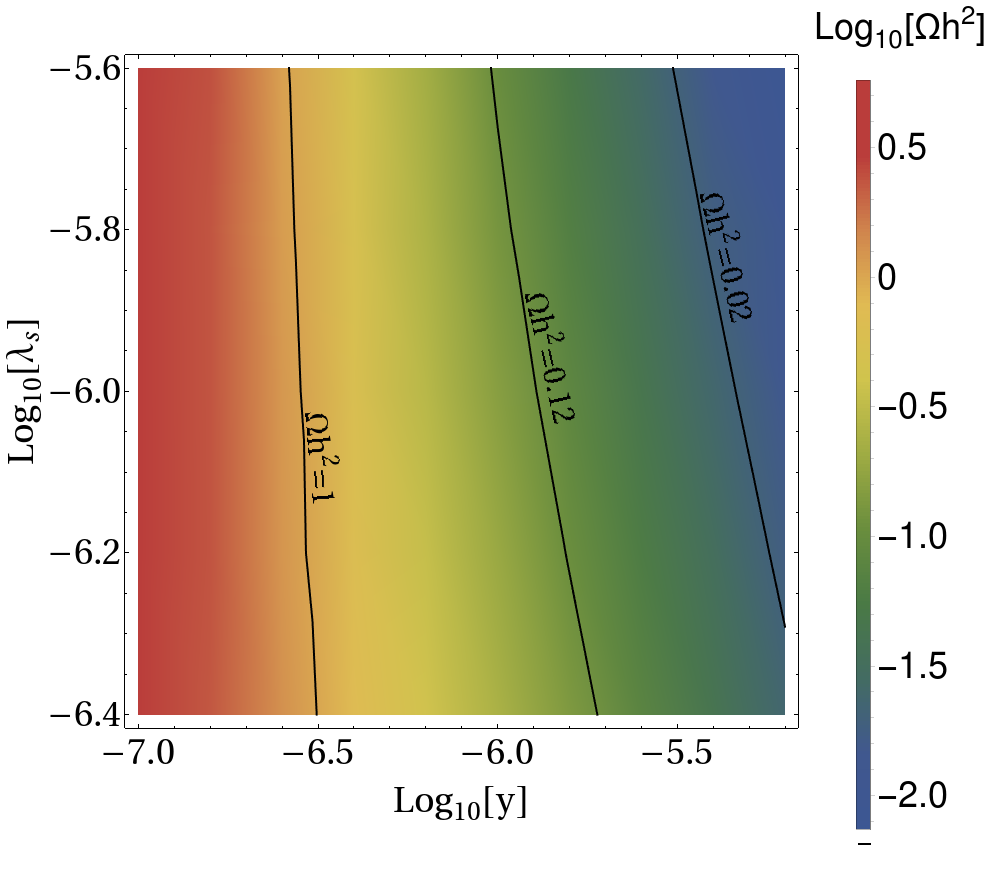} 
    \includegraphics[width=0.337\textwidth]{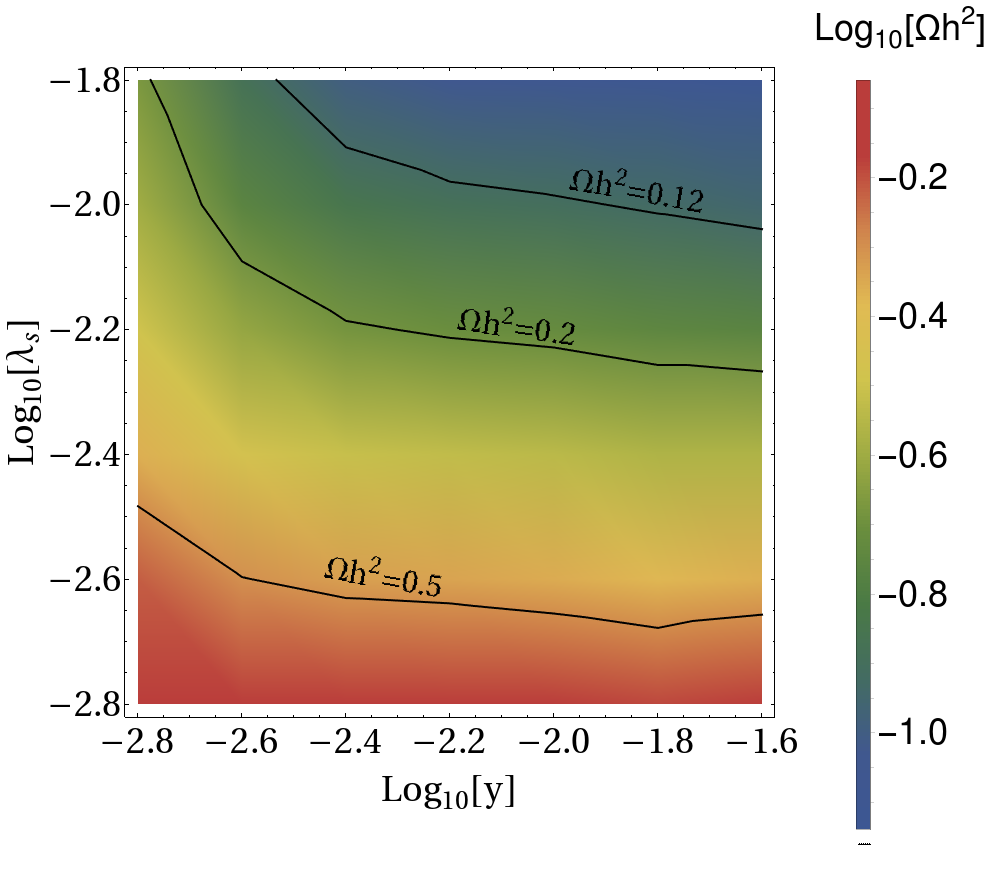} 
  \end{tabular}
\caption{Results of our numerical scans; in each of the panels we fix $y_M$. It reads $7\times 10^{-4}$ (left panel), $2\times 10^{-4}$ (middle) and $1.38\times 10^{-3}$ (right). The colors indicate the change of the relic density as a function of $y$ and $\lambda_S$ while the black lines indicate several specific values of $\Omega h^2$. In particular, each panel features $0.12$ line that matches observation. Red star in the left panel indicates our benchmark point (see \cref{fig:benchmark}).
}
\label{fig:scan}
\end{figure}

The VEVs, $v_S$ and $v_\sigma$, are obtained from the 
potential $V_\text{S+NJL}$ (\ref{VSNJL}), 
where $V_\text{NJL}$ is given in (\ref{eq:Vnjl}). 
Since their scale is fixed by $\Lambda_H$
(see also (\ref{NJL para})), we vary 
$\Lambda_H$ to obtain 
$v_S=(5/7) \times 10^{11}~\mbox{GeV}$.
Using these values for (\ref{VEVM}) 
we obtain the constituent mass
$\langle M (S,\sigma)\rangle=4.64\times 10^{10}$ GeV.
The DM and $S$ masses are calculated from 
(\ref{mS}) and (\ref{mDM}), respectively, 
where the wave function renormalization constant is $Z=3.98$.
The effective couplings $G_{\phi^2S^2}$ and $ G_{\phi^2S}$ 
are defined in (\ref{LphiS}) 
and can be computed from (\ref{Gp2S2}) and (\ref{Gp2S}),
where we use the approximated expressions  of the loop functions (see \cref{loop1} to \cref{loop5}).

In \cref{fig:benchmark} one can observe that across relevant temperature range $Y_{N_R}$ gradually approaches thermal equilibrium value. At smaller temperatures than those shown $N_R$ would decay into SM particles through lepton portal; it corresponds to the temperature range at which DM is already produced. Such decays, while irrelevant for DM production, can however play a role in producing required amount of baryon asymmetry of the Universe, as shown in 
Refs.\,\cite{Brdar:2019iem,Brivio:2019hrj}. $S$ abundance increases and reaches maximum slightly before it decays away. Since we take very tiny portal $\lambda_{HS}$ and in addition $S$ is typically lighter than $\phi$, the only available decay channel is into a pair of right-handed neutrinos (see \cref{S-decay}). Most importantly, we observe how DM candidate, $\phi$, neatly undergoes freeze-in; its abundance follows the characteristic curve for such non-thermal DM production. In particular, the strongest production occurs at the temperature of 
the order of DM mass scale, i.e. $T \simeq m_\text{DM}$. We have checked that the dominant and in fact only relevant production arises from the $\phi\phi\leftrightarrow N_R N_R$ process, leaving  $\phi\phi\leftrightarrow S S$ subdominant. Notice that the situation is very similar but still a little bit different with respect to the conventional freeze-in \cite{Hall:2009bx}: Namely, instead of producing DM through feeble interactions with SM particles that have a thermal abundance, in this case production occurs through
interactions with BSM particles that happen to have non-negligible abundance due to rather strong interactions with the thermal bath 
(decays and inverse decays of $N_R$).
We have also checked that the DM relic abundance $\Omega h^2$
does not depend on the starting temperature $T_0$ (at which all yields
are set equal to zero) as long as $T_0 > m_\text{DM}$ is satisfied.

In \cref{fig:scan} we present the results of the  parameter scans. As already elaborated in \cref{sec:model}, there are essentially only three relevant parameters (if $m_N$ is fixed):  $y_M$, $\lambda_S$ and $y$. In the left panel, we fix  $y_M$ to the value corresponding to our benchmark point; then $\Omega h^2=0.12$ 
is expected for values of $y$ and $\lambda_S$ given in \cref{input} (our benchmark point is shown as a red star in the left panel). In the middle and right panels we show the cases when $y_M$ is fixed to $2\times 10^{-4}$ and $1.38\times 10^{-3}$, respectively. In these two figures 
we demonstrate that for a given value of $y_M$, DM abundance can be relatively independent on one of the remaining two parameters: In the middle (right) panel $\Omega h^2$ is only mildly dependent on $\lambda_S$ ($y$). These are limiting cases, typically relic abundance is a non-trivial function of both parameters and clearly also strongly dependent on $y_M$; rather mild change of this parameter yields very different values of $y$ and $\lambda_S$ for which observed value of DM relic abundance is obtained.

Overall, we have shown that the successful production of DM is achievable in the vast portion of parameter space without employing any tuning.

\section{Summary and Conclusions}
\label{sec:summary}
\noindent
The neutrino option \cite{Brivio:2017dfq}
establishes a link between the heavy right-handed neutrinos, $N_R$,
and the electroweak scale. 
We have considered a scale invariant realization of the neutrino option
and  studied the possibility of incorporating DM into the  model.
In contrast to Refs. \cite{Brdar:2018vjq,Brdar:2018num}, where such realization
is achieved  perturbatively \`a la Coleman-Weinberg \cite{PhysRevD.7.1888}, 
we have coupled a strongly interacting 
QCD-like sector via a real SM singlet scalar $S$ to the SM with three
right-handed neutrinos introduced.
The scale invariance is dynamically broken 
by the chiral condensate  in the hidden sector,
and the scale generated in this way is transmitted via $S$ to the SM sector.
The dynamical chiral symmetry breaking in the hidden sector
produces NG bosons in the same way as in QCD.
 They are massive, because the Yukawa 
 coupling of $S$ with the hidden chiral fermions $\psi$
 explicitly breaks the chiral symmetry. Moreover, they are stable
 due to the unbroken  vector-like flavor symmetry
 ($SU(3)_V$ in our case)
 and therefore can be  DM candidates.
 The main task of this work has been
 to show that they can indeed be a realistic DM with the abundance that matches observations.

Unlike previously studied models of  similar kind 
\cite{Hur:2011sv,Heikinheimo:2013fta,Holthausen:2013ota,Hatanaka:2016rek,Kubo:2014ida,Ametani:2015jla}, in which
the quartic coupling $\lambda_{HS} S^2 H^\dag H$
is the portal for the transmitted scale, we have assumed that
the Yukawa coupling $y_M S N_R^T C N_R$ 
is the main portal, because 
the neutrino option scenario would be spoiled if 
 (the tree-level or bare) $\lambda_{HS}$ is not sufficiently suppressed.
This has a considerable influence on  the scale $\Lambda_H$ of the hidden sector:
Since the neutrino option works if 
the  right-handed neutrino mass, $m_N=y_M \langle S \rangle$,
is in the range $10^7 - 10^8$ GeV, $\Lambda_H$ 
can not be of $\mathcal{O}$(TeV) as is the case in Refs.\,\cite{Hur:2011sv,Heikinheimo:2013fta,Holthausen:2013ota,Hatanaka:2016rek,Kubo:2014ida,Ametani:2015jla}.
For $y_M \lsim 0.1$ we consequently have  $\langle S \rangle \gsim 10^8 - 10^9$ GeV,
which implies a large $\Lambda_H$.
More importantly, we have found that the DM mass $m_\text{DM}$
is  so large (several orders of magnitude above $\mathcal{O}$(TeV)) that
the freeze-out of DM is not achievable to obtain a consistent
DM relic abundance, and hence we have turned to 
the freeze-in mechanism \cite{Hall:2009bx}.
For the conventional freeze-in process, the DM should be sufficiently
disconnected with the thermal bath.
In our case, $\phi$ (DM), $S$ and $N_R$ interplay with each other
during the freeze-in  process in a non-trivial way, 
where, due to the assumption $\lambda_{HS} \sim 0$,
 $\phi$ and $S$ have only an indirect contact with
the SM via $N_R$ which has a direct contact with the SM through 
the Dirac-Yukawa  coupling.
Since $m_N$ and $y_\nu$ are basically fixed 
(by neutrino option), there are only three independent
parameters left: $y$ (for $S \bar{\psi} \psi$)  
and $y_M$  (for $S N_R^T C N_R$)   and  $\lambda_{S}$ (for $S^4$).
We have found that there exists a sufficient parameter space
 in which consistent DM relic abundance can be obtained.
We emphasize  that the DM in the model is successfully produced 
at temperature scale $T$ where the right-handed neutrinos are still stable,
i.e. $T \gsim m_N$.

Throughout the analysis we have assumed that the hidden sector,
described by a non-abelian gauge theory, is not thermalized.
This is because we have assumed that the hidden sector can be thermalized only
through a contact with the SM particles in the thermal bath;
if $S$ is not thermalized, the hidden sector, too, is not thermalized.
We have found that 
in the parameter space, in which the freeze-in mechanism for DM
works, the singlet $S$ is not thermalized.
That is, our hidden sector is cold and dark (``cold dark sector").
It would be an interesting question whether or not our DM scenario
works if the hidden sector is thermalized.
This is possible if, for instance, the hidden fermions have 
a non-vanishing SM $U(1)_Y$ hypercharge \cite{Kubo:2014ida,Ametani:2015jla}.
We may address this question in future publication.

We finally note that the lepton number asymmetry sufficient for the generation of observable baryon asymmetry of the Universe can be produced in right-handed neutrino decays as previously shown in the literature
 \cite{Brdar:2019iem,Brivio:2019hrj}  and such production is independent of DM production mechanism.  In combination with generation of light neutrino masses and absence of hierarchy problem (together with potential baryon asymmetry production from $N_R$ decays) we have hence demonstrated that the considered well-motivated model can successfully and simultaneously tackle some of the most relevant puzzles in high-energy physics.

 \section*{Acknowledgments}
 The work of M.~A. is supported in part by the Japan Society for the
Promotion of Sciences Grant-in-Aid for Scientific Research (Grant
Numbers 17K05412 and 20H00160). 
J.~K.~is partially supported by the Grant-in-Aid for Scientific Research (C) from the Japan Society for Promotion of Science (Grant No.19K03844).
\bibliographystyle{JHEP}
\bibliography{refs}

\end{document}